\documentclass[aps,prl,twocolumn,superscriptaddress,showpacs,amssymb,amsmath]{revtex4-1}
\usepackage{graphicx,multirow}

\begin{document}

\title{Selection rules for Raman-active electronic excitations in carbon nanotubes}

\author{Oleksiy Kashuba}
\affiliation{Institut f\"ur Theorie der Statistischen Physik, RWTH Aachen, 52056 Aachen, Germany}
\email{kashuba@physik.rwth-aachen.de}
\author{Vladimir I. Fal'ko}
\affiliation{Department of Physics, Lancaster University, Lancaster, LA1~4YB, UK}


\begin{abstract}
Raman measurements in carbon allotropes are generally associated with the exploration of the vibrational modes.
Here, we present a theory of the non-resonant inelastic light scattering accompanied by the excitations of intersubband electron-hole pairs in carbon nanotubes and predict the selection rules and polarization properties of the dominant intersubband Raman active modes.
\end{abstract}


\maketitle

Carbon nanotubes (CNTs) are one of the most exhaustively investigated allotropes of carbon~\cite{dresselhaus}.
In particular, CNTs have been extensively studied using optical spectroscopy: absorption of light~\cite{absorption1,absorption2,absorption3fluorescence2,absorption4,absorption5}, fluorescence~\cite{fluorescence1,absorption3fluorescence2,fluorescence3raman5,fluorescence4,fluorescence5,fluorescence6}, and inelastic (Raman) light scattering~\cite{raman1,raman2,raman3,raman4,fluorescence3raman5,raman6,book-r}.
Absorption and luminescence studies of CNTs addressed the electron-hole excitations in a semiconductor nanotubes with small radii, Raman spectroscopy was effectively employed for characterizing vibrational modes.
However, up to now, only one study~\cite{raman7el} has reported an observation interpreted as a double-resonant Raman inelastic scattering of light resulting in the creation of electronic excitation in CNTs.
Moreover, no experiment or theory has been reported, yet, on the non-resonant Raman scattering with the e-h excitations in the final state, similar to those observed in graphene.
In this Letter we offer a theory of inelastic light scattering in large-radius carbon nanotubes accompanied by the excitation of the low-energy electron-hole pairs in the final state of the non-resonant Raman process.

Theoretically, electronic properties of CNTs have a lot in common with those of graphene: a nanotube can be viewed as a rolled-up sheet of graphene.
Recently, Raman spectroscopy of electronic excitations in graphene was understood theoretically~\cite{erraman,erraman2} and, then, realized experimentally~\cite{erexp,*erexp2}.
In contrast to a non-relativistic plasma of free electrons where inelastic scattering of light is dominated by the second-order contact interaction of charge carriers instantaneously with two photons~\cite{platzmanwolff}, in graphene it is dominated by a two-step process consisting of sequential events of single-photon absorption and emission, with a virtual state between them.
This leads to the Raman spectrum $g(\omega)\propto\omega$ transforming into a pronounced structure of inter-Landau-level excitons in an external magnetic field, with the strongest resonances at the energies equal to twice the Landau level energy of the Dirac electron~\cite{erraman,erraman2,erexp,*erexp2}.

To compare, a structure of Raman spectra in carbon nanotubes should be determined by the quantization of electronic states into quasi-one-dimensional subbands, with the characteristic van Hove singularities near the subband edges.
Below, we determine the selection rules for the dominant Raman-active intersubband electron-hole excitations in CNTs, estimate their quantum efficiency, and find the relation between the polarizations of incoming and Raman-scattered photons.

\begin{figure}
\centering
\includegraphics[scale=.7]{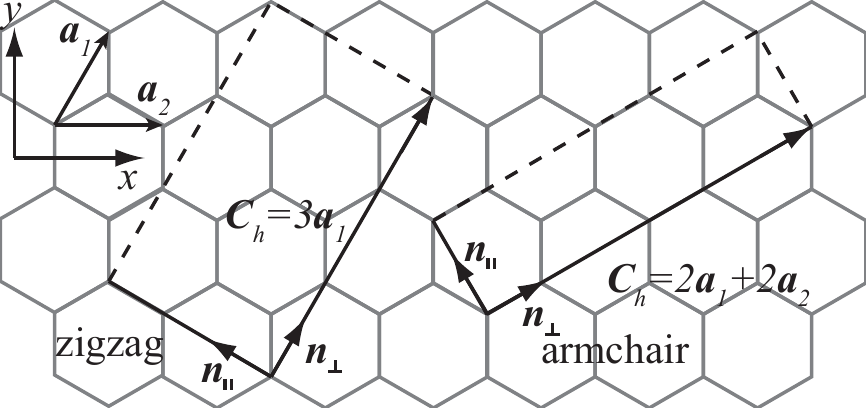}
\caption{Lattice structure for zigzag and armchair nanotube is shown.
Lattice vectors $\mathbf{a}_{1/2}$, $xy$ coordinate system for $\sigma_{x/y}$ matrices, chiral vector $\mathbf{C}_{h}$, and orthogonal unit vectors $\mathbf{n}_{||/\perp}$ are shown.
Nanotube is rolled up from a graphene sheet in such a way that lines parallel to $\mathbf{n}_{||}$ match.}
\label{fig:nanotube}
\end{figure}

\begin{figure}
\centering
\includegraphics[scale=.7]{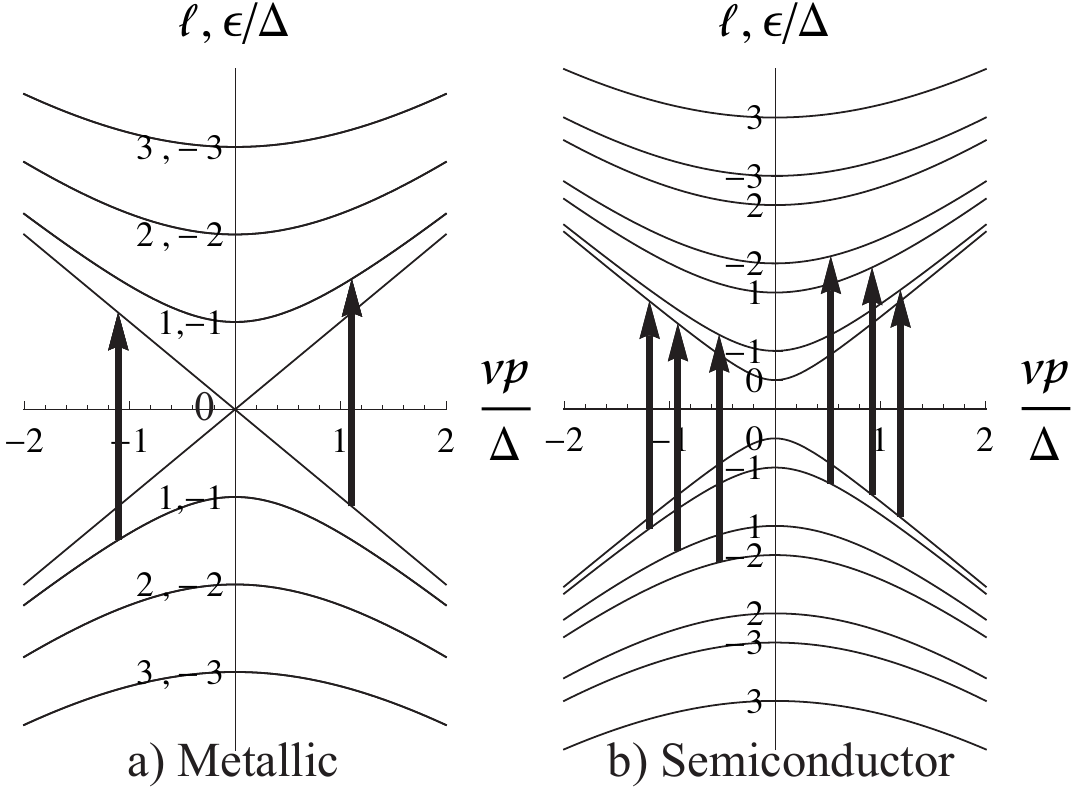}
\caption{Low-energy band structure for a) metallic (armchair; zigzag $n/3\in\mathbb{N}$ and chiral $(n+2m)/3\in\mathbb{N}$), and b) semiconductor (zigzag $n/3\notin\mathbb{N}$ and chiral $(n+2m)/3 \notin\mathbb{N}$) nanotubes.
Labels on $y$-axis denotes the numbers of subbands $l$, which are degenerate for metallic case.
For semiconductor case subband numbers are shown for $\delta=+^{1}\!\!/_{3}$ (for $\delta=-^{1}\!\!/_{3}$ the number's sign should be changed to the opposite).
The $y$-axis represents an electron momentum along the nanotube.
The transitions of an energy of $\omega=2.6\Delta$ are shown, which can change the subband number by $\pm1$.}
\label{fig:spectra}
\end{figure}

Using a close relation with graphene, illustrated in Fig.~\ref{fig:nanotube}, electron states in a carbon nanotube characterized by chirality vector $\mathbf{C}_{h}=n \mathbf{a}_{1} + m \mathbf{a}_{2}$~\cite{Note1} and diameter $D=a\frac{\sqrt{n^{2}+m^{2}+nm}}{\pi}$ can be described as plain waves,
\begin{equation}
\psi= e^{i ( p \mathbf{n}_{||} + (2\xi l/D) \mathbf{n}_{\perp})\mathbf{r}} \chi,
\end{equation}
with momenta $p \mathbf{n}_{||} + (2\xi l/D) \mathbf{n}_{\perp}$ counted from the $K$-points in the Brillouin zone of the honeycomb lattice.
Here we use the basis $\chi^\top=(\chi_{A}, \chi_{B})$ for the $K$ ($\xi=+$) and $\chi^\top=(\chi_{B}, \chi_{A})$ for $K'$ ($\xi=-$) valley, where $\chi_{A(B)}$ are components of the wave functions defined on the $A$($B$) sublattice of the honeycomb lattice, and $p$ is the electron valley momentum along the nanotube axis.
Integer $l$ stands for the quantum number characterizing the electron angular momentum around the nanotube, $\mathbf{n}_{\perp}=\mathbf{C}_{h}/|\mathbf{C}_{h}|$, and $\mathbf{n}_{||}\perp\mathbf{n}_{\perp}$.
Then, the CNT $2\times2$ Hamiltonian reads~\cite{dresselhaus}
\begin{equation}
H= \xi vp \mathbf{n}_{||} \cdot\boldsymbol{\sigma} + \Delta (l+\delta) \mathbf{n}_{\perp} \cdot\boldsymbol{\sigma},
\label{eq:hamiltonian}
\end{equation}
where $v$ is Dirac velocity in graphene, and $\Delta=2v\hbar/D$, and Pauli matrices $\boldsymbol{\sigma}=(\sigma^{x},\sigma^{y})$ act in the space of spinors $\chi$, and $\delta=\frac{1}{3}\mod_{3}(n+2m)$ is a minimal remainder of division of $n+2m$ by $3$.
The CTN spectrum (sketched in Fig.~\ref{fig:spectra}), with $s=\pm$ attributing states to the conduction ($s=+$) or valence ($s=-$) band,
\begin{equation}
\begin{split}
\epsilon_{slp}=s\sqrt{v^{2}p^{2}+\Delta^{2}(l+\delta)^{2}},
\quad
\chi_{\xi slp}=\frac{1}{\sqrt{2}}\binom{1}{e^{i\phi}},
\\
e^{i\phi}= \xi\frac{vp}{\epsilon_{slp}} (n_{||x}+in_{||y}) + \frac{\Delta (l+\delta)}{\epsilon_{slp}} (n_{\perp x}+in_{\perp y}),
\end{split}
\label{eq:wavefunctions}
\end{equation}
can be metallic ($\delta=0$), or semiconducting ($\delta=\pm^{1}\!/_{3}$).
All armchair nanotubes are metallic, while chiral and zigzag nanotubes can be both metallic or semiconducting, depending on their diameter~\cite{dresselhaus}.
All subbands in Eq.~\eqref{eq:wavefunctions} are valley and spin degenerate.

This relation between graphene and CNTs can be also exploited to describe the electron interaction with light,
\begin{equation}
H_{e-ph}= - \frac{ev}{c} \xi \boldsymbol{\sigma} \mathbf{A}.
\label{eq:emcoupling}
\end{equation}
Here, one takes into account that a projection of the vector potential $\overrightarrow{A}$ of the external electromagnetic field onto the unfolded sheet acquires periodic spatial dependence in its part perpendicular to the CNT axis,
\begin{equation}
\mathbf{A}= A_{||}\mathbf{n}_{||} + A_{\perp} \mathbf{n}_{\perp} \cos\frac{2\mathbf{n}_{\perp} \mathbf{r}}{D}.
\label{eq:avpcos}
\end{equation}
The latter feature sets the selection rules~\cite{ando} for the interband transitions excited by the electromagnetic field (see Table~\ref{tab:selrules}): $l\to l$ for $\overrightarrow{A}$ polarized along the CNT axis ($A_{\perp}=0$) and $l\to l\pm1$ for $\overrightarrow{A}$ polarized perpendicular to the CNT axis ($A_{||}=0$).
These selection rules determine the dominant lines in the absorption~\cite{absorption2,fluorescence1,fluorescence5} and luminescence~\cite{fluorescence1,fluorescence3raman5,fluorescence5} spectra of long CNTs, where the processes with $\mathbf{A}||\mathbf{n}_{||}$ may be additionally enhanced by the antenna effect.

The electron-photon interaction in Eqs.~(\ref{eq:emcoupling},\ref{eq:avpcos}) can be used to establish the selection rules for the inelastic scattering of photons from a large-radius CNT, with an electron-hole pair left in the final state at the excitation energy $\omega\ll\Omega$, where $\Omega$ is the energy of incoming photon.
In 2D graphene, the main contribution to the amplitude of inelastic light scattering is given by two Feynman diagrams,
\begin{equation}
R_{11} =
\raisebox{-.3in}{\includegraphics[height=.7in]{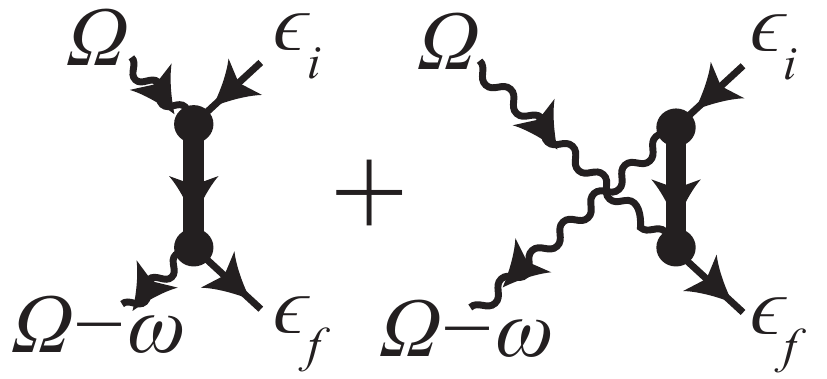}}\quad,
\label{eq:diagrams}
\end{equation}
describing sequential absorption and emission of individual photons, one corresponding to the absorption event preceding emission with a large excess of energy $\Omega$ in the intermediate state, and the other with emission preceding absorption and energy deficit $-\Omega$ in the intermediate state.
In these Feynman diagrams, thin straight lines identify electrons in the ``in'' and ``out'' states with the energies $\epsilon_{i}$ and $\epsilon_{f}$, wavy lines --- absorbed and emitted field $\mathbf{A}$ and $\tilde{\mathbf{A}}$.
The thick line in $R_{11}$ is the electron propagator in the intermediate state, $G = \sum_{n}\frac{|n\rangle \langle n|}{\epsilon_{i}+\Omega-\epsilon_{n}}$ for the first diagram and $G = \sum_{n}\frac{|n\rangle \langle n|}{\epsilon_{f}-\Omega-\epsilon_{n}}$ for the second, where $\omega=\epsilon_{f}-\epsilon_{i}$ is the Raman shift, and the sum is taken over the intermediate electron states $|n\rangle$ with energies $\epsilon_{n}$.
For the case of $\Omega\gg\Delta,\omega$ studied here, one can approximate $G\approx \pm \Omega^{-1}$.
This approximation enables us to simplify the expression for the effective interaction of electrons with the pair of photons to the form
\begin{multline}
R_{11} \approx \frac{e^{2}v^{2}}{c^{2}}\left(\frac{1}{\Omega}(\boldsymbol{\sigma}\mathbf{A})(\boldsymbol{\sigma}\mathbf{\tilde{A}})+\frac{1}{-\Omega}(\boldsymbol{\sigma}\mathbf{\tilde{A}})(\boldsymbol{\sigma}\mathbf{A})\right)
=\\=
\frac{e^{2}\hbar^{2}}{2\Omega}(e_{||}\tilde{e}_{\perp}-e_{\perp}\tilde{e}_{||}) \cos \left(\frac{2\mathbf{n}_{\perp}\mathbf{r}}{D} \right) \frac{2i\sigma ^{z}}{\Omega},
\label{eq:R}
\end{multline}
where $e_{||(\perp)}$ and $\tilde{e}_{||(\perp)}$ is polarization of the incident and scattered light parallel (perpendicular) to the CNT axis.
For the dominant~\cite{Note2} Raman active mode (see Table~\ref{tab:selrules}) this determines the selection rule 
\begin{equation}
l\to l\pm1 \quad\text{with} \quad p=p',
\end{equation}
and probability
\begin{align}
\label{eq:W}
W &= \frac{(e\hbar v)^{4}}{\Omega^{4}}(e_{||}\tilde{e}_{\perp}-e_{\perp}\tilde{e}_{||})^{2}
\times\\&\times
\frac{2\pi}{\hbar}\sum_{s,s',l,\pm,\xi}\int\frac{d p}{2\pi\hbar}
\left| \chi_{\xi,s', l\pm1}^{+} \sigma^{z} \chi_{\xi,s,l}\right|^{2}
\times\nonumber\\&\times
f(\epsilon_{s,l,p}) (1-f(\epsilon_{s',l\pm1,p})) \delta(\epsilon_{s,l,p}-\epsilon_{s',l\pm1,p}+\omega).
\nonumber
\end{align}
Here, $f(\epsilon)=1/(e^{\frac{\epsilon-\mu}{T}}+1)$ is the occupation number of electron states, and
\begin{multline}
\left| \chi_{\xi,s', l\pm1,p}^{+} \sigma^{z} \chi_{\xi,s,l,p}\right|^{2} = \frac{1}{2}-\frac{ss'}{2}\times\\\times\frac{\Delta^{2}(l\pm1+\delta)(l+\delta)+v^{2}p^{2}}{\sqrt{\Delta^{2}(l\pm1+\delta)^{2}+v^{2}p^{2}}\sqrt{\Delta^{2}(l+\delta)^{2}+v^{2}p^{2}}},
\label{eq:melement}
\end{multline}
is a projector dependent on whether the transition is interband ($s=s'$) or intraband ($s=-s'$), on the value of the subband index $l$, and on the momentum $p$ in the initial state of the photoexcited electron.
For the interband ($s'=-s$) transitions, Eq.~\eqref{eq:melement} gives the values of the matrix elements which are close to $1$ for both $l\gg1$ and $vp\gg\Delta$.
Also, for $l\sim1$ and $vp\lesssim\Delta$ we find that $|\chi|^{2}\lesssim1$, except for the transitions $-1\to0$ ($\delta=+^{1}\!/_{3}$), and $0\to1$ ($\delta=-^{1}\!/_{3}$) in a semiconductor nanotube, where $|\chi|^{2}=0$ when $p=0$.
For the intraband ($s'=s$) transitions behavior of the amplitude in Eq.~\eqref{eq:melement} is quite different: $|\chi|^{2}$ vanishes both when $p=0$ and when $l, vp/\Delta\gg1$, with $|\chi|^{2}\sim1$ at $l\sim vp/\Delta\sim1$.
Exception, once again, is given by the transitions $0\to-1$ and $-1\to0$ in a semiconductor CNT, which show behavior characteristic for the interband transitions.

The ratio between the power of scattered and incoming light, spectral density $g=P_{out}/P_{in}$ is given by
\begin{multline}
g(\omega)=
\frac{\pi \nu \Omega^{2} \Delta}{(hc)^{4} v \sin\theta}\, W
=\\=
\frac{1}{16\pi} \left(\frac{e^{2}}{h c}\frac{v}{c}\right)^{2}
\frac{\left(e_{||}\tilde{e}_{\perp}-e_{\perp}\tilde{e}_{||}\right)^{2}}{\sin\theta}
\nu\frac{\Delta}{\Omega^{2}} F(\omega/\Delta).
\end{multline}
Here, $\nu$ characterizes areal density of CNTs with given diameter and chirality, and factors
\begin{equation}
F(\alpha)=
\sum_{\pm,l\geqslant0}
\sqrt{\left|\frac{\alpha^{2}-1}{\alpha^{2}-\eta_{\pm}^{2}}\right|},
\qquad
\eta_{\pm}=2l+1\pm 2\delta,
\label{eq:Ja}
\end{equation}
reflect van Hove singularities of the quasi-1D subbands in the nanotube.
The rules which determine the limits of the sum in Eq.~\eqref{eq:Ja} are as follows: \\ (a) if $\alpha>1$, then for the interband processes 
\begin{equation*}
\eta_{\pm}<\alpha, \quad
\begin{cases}
2\alpha\frac{\mu}{\Delta}-\eta_{\pm}<\alpha^{2}, &\text{for $\pm l \to \pm (l+1)$}
\\
2\alpha\frac{\mu}{\Delta}+\eta_{\pm}<\alpha^{2}, &\text{for $\pm (l+1) \to \pm l$}
\end{cases};
\end{equation*}
(b) if $\alpha<1$, then only intraband processes $\pm l \to \pm (l+1)$ are allowed (we assume $\mu>0$), with
\begin{equation*}
\eta_{\pm}>\alpha, \quad \left|2\alpha\frac{\mu}{\Delta}-\eta_{\pm}\right|<\alpha^{2}.
\end{equation*}
Since $F(\alpha\gg1)\approx \pi \alpha$, asymptotically, for $\omega\gg\Delta$, $g\propto \omega/\Omega^{2}$, as in 2D graphene~\cite{erraman}.
Besides the dominant Raman-active electron-hole excitations, there are weaker processes~\cite{Note2} listed in Table~\ref{tab:selrules}.

\begin{table}
\centering
\begin{tabular}{c|c|c|c|}
& $l\to l$ & $l\to l\pm1$ & $l\to l\pm2$ \\
\hline
\hline
\multirow{2}{40mm}{\parbox{40mm}{Absorption (emission) \\ active mode}}
& $||$ & 0 & 0 \\
\cline{2-4}
& 0 & $\perp$ & 0 \\
\hline
\hline
Dominant Raman mode & 0 & $\perp \leftrightarrow ||$ & 0 \\
\hline
\multirow{3}{40mm}{\parbox{40mm}{Weak Raman \\ active modes~\cite{Note2}}}
& 0 & $\perp \leftrightarrow||$ & 0 \\
\cline{2-4}
& $||\to||$ & 0 & 0 \\
\cline{2-4}
& $\perp\to\perp$ & 0 & $\perp\to\perp$ \\
\hline
\end{tabular}
\caption{Comparison of selection rules and polarization properties for absorption/emission and Raman processes.}
\label{tab:selrules}
\end{table}

\begin{figure}
\centering
\includegraphics[scale=.6]{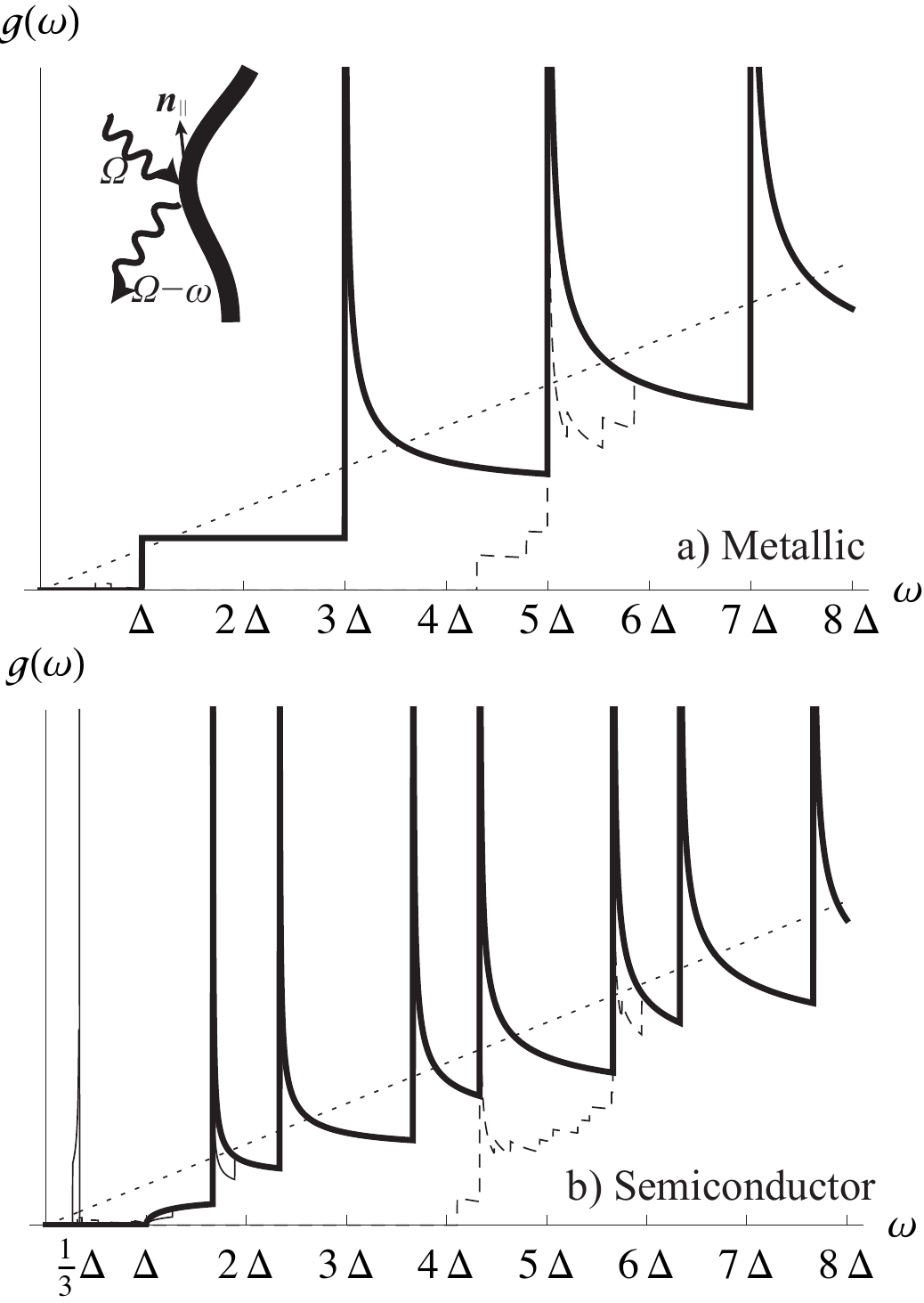}
\caption{The predicted Raman spectra of carbon nanotubes.
Inset shows a schematic image of the experiment with incoming light at frequency $\Omega$, emitted light with frequency $\Omega-\omega$, and local direction of the nanotube axis, $\mathbf{n}_{||}$.
Plots show Raman spectra of a) metallic (peaks at $\omega = (2l+1)\Delta$, $l\in\mathbb{N}$) and b)semiconductor (peaks at $\omega = (2l\pm^{1}\!\!/_{3})\Delta$) nanotubes.
Thick line is for undoped case; dashed is for doped CNT with Fermi energy $\mu=2.5\Delta$; dotted line shows the linear spectra for undoped graphene.
Thin line represents Raman spectrum of a semiconducting nanotube with $\mu=0.5\Delta$, in which additional peak at $\omega=\Delta/3$ appears.}
\label{fig:ntrspectra}
\end{figure}

Several examples of the resulting Raman spectra of electronic excitations in metallic and semiconducting CNTs are plotted in Fig.~\ref{fig:ntrspectra}(a,b).
For each of the two nanotube types we show the spectra for undoped (solid line) and doped (dashed line) CNTs.
For each value of the Raman shift $\omega$ spectral density $g(\omega)$ plotted in Fig.~\ref{fig:ntrspectra} is composed from the contributions of several intersubband transitions with characteristic van Hove singularities.
There is one exception from this rule: when the transitions $0\leftrightarrow-1$ turn on in undoped CNTs with $\delta=+^{1}\!/_{3}$, Raman intensity experiences a continuous increase from the threshold at $\omega=\Delta$, which is due to a peculiar $p$-dependence of the transition amplitude in Eq.~\eqref{eq:melement}.
Also, in metallic nanotubes, $g(\omega)$ experiences a simple jump (Fig.~\ref{fig:ntrspectra}(b)) when the lowest intersubband transitions $0\leftrightarrow\pm1$ turn on at $\omega=\Delta$, reflecting the linear dispersion of electrons in the subband with $l=0$.

Doping of nanotubes allows for some intraband transitions with $\omega<\Delta$.
Small doping with Fermi energy $\Delta/3<\mu<2\Delta/3$, allows for intraband transitions with the amplitude $\sim1$, including the most prominent line $0\leftrightarrow-1$ shown in Fig.~\ref{fig:ntrspectra}(b), the only one which results in the intraband van Hove singularity at $\omega=\Delta/3$.
For higher doping, $\mu>\Delta$, Raman spectrum is strongly suppressed at the energies $\omega<\Delta$ due to the small Raman amplitudes of the intraband processes described by Eq.~\eqref{eq:melement} and  Pauli blocking of the interband processes with $\Delta<\omega<2\mu-\Delta$.

To summarize, we determined selection rules and calculated the spectral density of electronic excitations in the non-resonant Raman spectrum of a single-wall carbon nanotube of a given chirality.
We found that the strongest Raman-active mode corresponds to the $l\to l\pm1$ intersubband transition~\cite{Note2}, and a characteristic CNT spectrum we obtained (see, e.g.~in Fig.~\ref{fig:ntrspectra}) features van Hove singularities corresponding to the electron excitations from/to the top/bottom of the corresponding nanotube subbands.
These selection rules are specific for the non-resonant Raman scattering in large-radius nanotubes, where the e-h excitation energies are much less than the in/out photon energies.

The authors thank M.~Potemski, T.~Heinz, I.~Aleiner, and J.~K\"urti for useful discussion and EPSRC and Royal Society for financial support.

\end{document}